\documentclass{article}
\usepackage{spconf,amsmath,graphicx}
\usepackage{amssymb}
\usepackage{booktabs}
\usepackage{bm,amsfonts}
\usepackage{threeparttable}
\usepackage{float}
\def\x{{\mathbf x}}

\def\x{{\bm x}}

\def\bpi{{\bm \pi}}
\def\B{{\mathcal B}}
\def\bt{{\bm \theta}}

\usepackage{hyperref}
\title{Exploiting Single-Channel Speech For Multi-channel End-to-end Speech Recognition}

\name{Keyu An, Zhijian Ou$^{\dagger}$\thanks{$\dagger$ Corresponding author.}}
\address{Speech Processing and Machine Intelligence (SPMI) Lab, Tsinghua University, China}

\begin{document}
\ninept
\maketitle
\begin{abstract}
Recently, the end-to-end training approach for neural beamformer-supported multi-channel ASR has shown its effectiveness in multi-channel speech recognition. However, the integration of multiple modules makes it more difficult to perform end-to-end training, particularly given that the multi-channel speech corpus recorded in real environments with a sizeable data scale is relatively limited. In this paper, we explore the usage of single-channel data to improve the multi-channel end-to-end speech recognition system. Specifically, we design three schemes to exploit the single-channel data, namely pre-training, data scheduling, and data simulation. Extensive experiments on CHiME4 and AISHELL-4 datasets demonstrate that all three methods improve the multi-channel end-to-end training stability and speech recognition performance, while the data scheduling approach keeps a much simpler pipeline (vs. pre-training) and less computation cost (vs. data simulation). Moreover, we give a thorough analysis of our systems, including how the performance is affected by the choice of front-end, the data augmentation, training strategy, and single-channel data size.
\end{abstract}
\begin{keywords}
multi-channel end-to-end ASR, neural beamformer, pre-training, data scheduling, data simulation.
\end{keywords}
\section{Introduction}
\label{sec:intro}
With the advances in deep neural networks, significant progress has been made in automatic speech recognition (ASR). However, speech recognition in far-field scenarios is still a challenging task~\cite{chime4,chime6}. To be specific, Fu~\emph{et al.}~\cite{aishell4}  report an over 30\% character error rate (CER) on AISHELL-4, which is recorded in distant-talking conference scenario, and Watanabe~\emph{et al.}~\cite{chime6}  report an over 50\% word error rate (WER) on chime-6, which is recorded in everyday home environments with distant microphones. 

Leveraging multi-channel signals has been shown to improve speech recognition performance in far-field scenarios~\cite{Beamnet_heymann,Multich-e2e}. The classical approach for multi-channel speech process is beamforming. Traditionally, the beamformer is optimized under the criteria such as the maximization of the output SNR or the Minimum Variance Distortionless Response (MVDR). After that, the enhanced output produced by the beamformer is processed with a single-channel ASR back-end. Despite its competitive results on a wide range of benchmarks~\cite{chime4,aishell4,USTC-iFlytek}, a drawback suffered by the classical approach is that the beamforming front-end and the ASR back-end are optimized separately under different criteria, and the complementary correlation between the two sub-tasks is ignored. To address this issue,  there have been research efforts to build multi-channel speech recognition system with a unified network~\cite{Beamnet_heymann,Multich-e2e,joint_tencent,3dcnn}. In such models, gradients from the acoustic model can be propagated to the front-end, which is typically a neural beamformer. Experiments show that joint optimization of the front-end and back-end reduces the mismatch of a combined system, and helps to improve the recognition performance~\cite{Beamnet_heymann,joint_tencent}.

Nevertheless, the integration of multiple modules significantly increases the difficulty of optimization, especially when no intermediate criterion is available for the system. For example, Heymann \emph{et al.}~\cite{Beamnet_heymann} find that training the integrated system from scratch leads to sub-optimal results, 
and in~\cite{mimo}, special strategies such as data scheduling and curriculum learning are adopted to facilitate the training process. Moreover, the access of multi-channel labeled data recorded in real scenarios is relatively more difficult than the single-channel data, which increases the difficulty of building a promising multi-channel end-to-end speech recognition model.

In this paper, we explore three ways to exploit single-channel speech for multi-channel end-to-end speech recognition, namely pre-training, data scheduling, and data simulation. On top of the previous research, the main contributions of this paper are: 

1) We extend the existing CTC-CRF-based speech~\cite{CRF_IC19,CAT} recognition
framework, by integrating multi-channel speech enhancement components into the unified framework. Aided by data augmentation and language modeling, our model obtains competitive results on CHiME4 and AISHELL-4 benchmarks without the use of model combination.

2) We compare the effectiveness of three strategies to utilize single-channel speech to improve the multi-channel end-to-end speech recognition system.

3) We give a thorough analysis of the factors that would affect the performance of the multi-channel end-to-end speech recognition system, including the choice of the front-end, data augmentation, the training strategy, and the amount of the single-channel speech data.

The rest of the paper is organized as follows. 
Section 2 outlines related work, especially focusing on the end-to-end training of multi-channel speech recognition system. 
Section 3 describes the mask-based MVDR neural beamformer as the front-end. 
Section 4 describes CTC-CRF based AM as the back-end and Section 5 focuses on describing the overall processing pipeline. 
In Section 6, the methods to exploit single-channel data are introduced in detail.
Section 7 presents and analyzes the results primarily on the CHiME4 and AISHELL-4 tasks. 
Section 8 presents conclusions.

\section{Related Work} 
\label{sec:related-works}
\subsection{End-to-end training of multi-channel ASR system}
Many works have focused on improving the recognition accuracy of far-field speech by leveraging multi-channel signals ~\cite{chime4,chime6,Beamnet_heymann}. 
The traditional approach to multi-channel ASR combines all the available channels by beamforming ~\cite{beamformit, mask_heymann} and then processes the resulting enhanced single with a separately trained single-channel AM. 
However, this approach has a few drawbacks~\cite{Beamnet_heymann}. Crucially, the beamformer and the acoustic model are optimized separately. Thus, the information from the acoustic model can not be utilized to improve the beamformer. Moreover, for neural beamformer training, clean speech data is required as the target, which is much more difficult to collect than noisy data in many scenarios. To overcome the above drawbacks, end-to-end training of multi-channel ASR system has been explored, which can be categorized into two approaches: multi-channel acoustic model without an explicit beamformer and neural beamformer based approach. The two approaches are described as follows.

The first approach is the multi-channel acoustic model without an explicit beamformer, where the neural network is seen as a replacement for conventional beamformer. For example, ~\cite{concatenate} simply concatenates multiple channel features and improves far-field ASR performance over single-channel input. However, direct concatenation makes it difficult for the neural network to learn the complex relationships between microphones due to the large feature dimensionality and numerous independent neural parameters ~\cite{concatenate}. To address it, several neural operations were proposed. For example, \cite{cnn_distant} proposed to use shared neural network weights across different channels, and a cross-channel pooling to combine the information from multiple channels. 3-D CNN~\cite{3dcnn} processes time, frequency and channel dimensions of the input spectrogram with a three-dimensional convolutional neural network, and~\cite{qnn} uses quaternion neural networks to model the complex inter- and intra- channel dependencies. In general, these methods require complex model architecture designs, and the models depend on the microphone configurations. Therefore, once the number and order of microphone channels are changed, the neural network has to be reconfigured and retrained.

In neural beamformer based approach,  the neural beamformer is cast as a differentiable component to allow joint optimization of the multi-channel speech enhancement with the ASR criterion. This approach can be further categorized into two types:  1)  mask estimation method, in which the neural network is used to estimate the time-frequency
masks, which are used to compute the statistics of speech and noise. Then, using these statistics, the filter coefficients are computed within the framework of constrained optimization problem of noise reduction, such as minimum variance distortionless response (MVDR) and generalized eigenvalue (GEV) ~\cite{Beamnet_heymann,Multich-e2e}. 2)  filter estimation method, in which the neural network is used to estimate the filter coefficients directly~\cite{deep_beamforming_xiao,adaptive_beamforming}. Compared with mask estimation method, filter estimation method lacks restriction to the filter coefficients estimation, and is usually more difficult to optimize due to the high flexibility of the estimated filters~\cite{Multich-e2e,deep_beamforming_xiao}. Moreover, similar to the multi-channel acoustic model without an explicit beamformer, filter estimation network is dependent on the microphone configurations. In light of the above trends, we choose mask based neural beamformer as our front-end.
\subsection{The optimization issues of multi-channel end-to-end system}
Despite the promising results, a lot of researches have pointed out that it becomes more difficult to train the multi-channel end-to-end system, which is composed of several modules~\cite{Beamnet_heymann,mimo,branches}. Specifically, performing straightforward end-to-end optimization of such a system can cause training failure~\cite{unified}, or lead to sub-optimal results~\cite{Beamnet_heymann}. To address this issue, some previous works suggest initializing the front-end and back-end with the respective pre-trained model~\cite{Beamnet_heymann,unified}. In the heterogeneous-input model that has both single-channel and multi-channel input branches~\cite{branches}, a single-channel
AM is trained first, then a multi-channel AM is trained starting from the single-channel AM with a randomly initialized multi-channel input branch. Different from our work, these works did not focus on the use of external single-channel data. In ~\cite{mimo},  curriculum learning is adopted to make the model learn firstly from easy samples and then more difficult ones. Moreover, data scheduling, which means every batch is randomly chosen either from the multi-channel set or from the single-channel
set during training, is proposed to regularize the training process. However, the training strategy (e.g. the choice of optimizers when the data comes from two sources) and the effect of using single-channel data in different scales are not thoroughly discussed in the original paper.

\section{Mask-based MVDR Neural Beamformer}
We adopt state-of-the-art MVDR neural beamformer ~\cite{Multich-e2e,MVDR} (illustrated in Fig. \ref{fig:res}(a) as the front-end. The following subsections mainly describe modules in the beamformer in detail.

\subsection{MVDR formulation}
MVDR reduces the noise and recovers the signal component by applying a linear filter to the overall observation vector:
\begin{equation} \label{eq:beforming} 
\hat{x}(t,f) = \sum^{C}_{c=1} h(f,c) \times x(t,f,c)
\end{equation}
where $ x(t,f,c)\in \mathbb{C} $  is a short-time
Fourier transform (STFT) coefficient at a time-frequency bin $(t, f)$ of the noisy signal at microphone $c$.  $ \hat{x}(t,f)  \in \mathbb{C} $
is the enhanced STFT coefficient, and $C$ is the numbers of microphones. According to the MVDR formulation ~\cite{MVDR},  the time-invariant filter coefficient ${\rm \textbf h}(f) = \{h(f,c)\}_{c=1}^C \in \mathbb{C}^C $ can be obtained by
$$
{\rm \textbf h}(f) = \frac{{{\bf \Phi_{NN}^{-1}}(f)}{ \bf \Phi_{SS}}(f)} {{\rm tr} \{ {{\bf \Phi_{NN}^{-1}} (f)}{ \bf \Phi_{SS}}(f)\}} { \rm \textbf u} 
$$
where ${\bf \Phi_{SS}} (f) \in \mathbb{C}^{C \times C} $ and ${\bf \Phi_{NN}} (f) \in \mathbb{C}^{C \times C} $ are the cross-channel power spectral density (PSD) matrices (also known as spatial covariance matrices) for speech and noise signals respectively.
${ \rm \textbf u} \in \mathbb{C}^C $ vector is a one-hot vector,  indexing the reference microphone. In practice, the reference microphone can be selected by principal component analysis ~\cite{mask_heymann}, or neural network based methods ~\cite{Multich-e2e}. $ \rm tr\{  \cdot  \} $  is the matrix trace operation.
\subsection{Cross-channel PSD estimation}

In our model, ${\bf \Phi_{SS}} (f) $ and ${\bf \Phi_{NN}} (f)  $ are estimated with a mask-based approach~\cite{mask_heymann}:
$$
{\bf \Phi_{SS}} (f) = \frac{1}{\sum_{t=1}^T m_{S}(t,f)} \sum_{t=1}^T m_{S}(t,f) {\rm \textbf x}(t,f) {\rm \textbf x^{\dagger}}(t,f)
$$
$$
{\bf \Phi_{NN}} (f) = \frac{1}{\sum_{t=1}^T m_{N}(t,f)} \sum_{t=1}^T m_{N}(t,f) {\rm \textbf x}(t,f) {\rm \textbf x^{\dagger}}(t,f)
$$
where ${\rm \textbf x}(t,f) = \{x(t,f,c)\}_{c=1}^C \in \mathbb{C}^C $, and $T$ is the length of the input features. $\dagger$ represents the conjugate transpose. $ m_{S}(t,f) \in [0,1] $ and  $ m_{N}(t,f) \in [0,1] $ are the time-frequency masks for speech and noise respectively \footnote{Note that in case of noise-aware training, $ m_{S}(t,f)$ and $m_{N}(t,f) $ are estimated separately (section \ref{sec:mask-estimation}). Thus, $ m_{S}(t,f) + m_{N}(t,f) $ is not necessarily equal to 1.}, and condensed from the masks
for all channels using a mean operation ~\cite{Multich-e2e}:
$$
m_{S}(t,f) = \frac{1}{C} \sum_{c=1}^{C} m_{S}(t,f,c)
$$
$$
m_{N}(t,f) = \frac{1}{C} \sum_{c=1}^{C} m_{N}(t,f,c)
$$
\subsection{Mask estimation}
\label{sec:mask-estimation}
In the mask estimation network approach, time-frequency masks for speech and noise are estimated separately with two neural networks:
$$
{\rm \bf  Z_{S}}(c) = {\rm BLSTM}(|{\rm \bf  X}(c)|)
$$
$$
{\rm \bf  m_{S}}(c) = {\rm sigmoid}({\bf W_S Z_{S}}(c) + {\bf b_S})
$$
$$
{\rm \bf  Z_{N}}(c) = {\rm BLSTM}(|{\rm \bf  X}(c)|)
$$
$$
{\rm \bf  m_{N}}(c) = {\rm sigmoid}({\bf W_N Z_{N}}(c) + {\bf b_N})
$$
where ${\rm \bf  Z_{S}}(c) = \{{\rm \bf z_{S}}(t,c) \in \mathbb{R}^{D_H} | t=1,...,T\}$, ${\rm \bf  Z_{N}}(c) = \{{\rm \bf z_{N}}(t,c) \in \mathbb{R}^{D_H} | t=1,...,T\}$,  and $D_H$ is the output dimension of the BLSTM. ${\rm \bf  m_{S}}(c) = \{{\rm  m_{S}}(t,c) \in \mathbb{R}^{F} | t=1,...,T\}$, ${\rm \bf  m_{N}}(c) = \{{\rm  m_{N}}(t,c) \in \mathbb{R}^{F} | t=1,...,T\}$, 
$|{\rm \bf  X}(c)| = \{|{\rm \bf x}(t,c)| \in \mathbb{R}^{F} | t=1,...,T\}$. $F$ is the dimension of STFT features. $|{\rm \bf  x}(t,c)| = \{| x(t,f,c)| \in \mathbb{R} | f=1,...,F\}$  is calculated as the norm of the complex STFT coefficients:
$$
| x(t,f,c)| = \sqrt{\Re^2( x(t,f,c)) + \Im^2( x(t,f,c))}
$$

\section{CTC-CRF based Speech Recognition}
This section explains CTC-CRF based speech recognizer ~\cite{CRF_IC19,CAT}.
Consider discriminative training with the loss defined by conditional maximum likelihood ~\cite{CRF_IC19}:
\begin{equation} \label{eq:crf-obj1}
\mathcal{L}(\bt) = - \log p_{\bt}(\bm{l} |\x)
\end{equation}
where $\x \triangleq x(1),\cdots\, x(T)$ is the speech feature sequence and $\bm{l}  \triangleq l_1, \cdots\, l_L$ is the label (phone, character, word-piece and etc) sequence, and $\bt$ is the model parameter.
Note that $\x$ and $\bm{l}$ are in different lengths and usually not aligned.
To handle this, a hidden state sequence $\bpi \triangleq \pi_1,\cdots\,\pi_T$ is introduced, and 
the label sequence is obtained by the mapping $ \B$ that removes consecutive repetitive labels and blanks in the hidden state sequence. Thus, the posteriori of $\bm{l} $ is defined through the posteriori of $\bpi$ as follows:
\begin{equation} \label{eq:post-l} 
	p_{\bt}(\bm{l} | \x) = \sum_{\bpi \in \mathcal{B}^{-1}(\bm{l} )} p_{\bt}(\bpi | \x)
\end{equation}
And the posteriori of $\bpi$ is further defined by a CRF:
\begin{equation} \label{eq:post-pi}
p_{\bt}(\bpi|\x) = \frac{\exp(\phi_{\bt}(\bpi, \x))}{\sum_{\bpi'}{\exp(\phi_{\bt}({\bpi', \x}))}}
\end{equation}
Here $\phi_{\bt}(\bpi, \x)$ denotes the potential function of the CRF, defined as:
\begin{displaymath}
\phi_{\bt}(\bpi, \x) = \log p(\bm{l} )+ \sum_{t=1}^{T} \log p_{\bt}(\pi_t|\x)
\end{displaymath}
where $\bm{l}  = \B(\bpi)$. $\sum_{t=1}^{T} \log p_{\bt}(\pi_t|\x)$ defines the node potential, calculated from the bottom AM DNN.
$\log p(\bm{l} )$ defines the edge potential, realized by an n-gram LM of labels.

Combining Eq. (\ref{eq:crf-obj1})-(\ref{eq:post-pi}) yields the sequence-level loss used in CTC-CRF:
\begin{equation} \label{eq:crf-obj2}
\mathcal{L}(\bt) = - \log \frac{  \sum_{\bpi \in \mathcal{B}^{-1}(\bm{l} )} \exp(\phi_{\bt}(\bpi, \x))}{\sum_{\bpi'}{\exp(\phi_{\bt}({\bpi', \x}))}}
\end{equation}

Remarkably, regular CTC suffers from the conditional independence between the states in $\bpi$. In contrast, by incorporating $\log p(\bm{l} )$ into the potential function in CTC-CRF, this drawback is naturally avoided. It has been shown that CTC-CRF outperforms regular CTC consistently on a wide range of benchmarks, and is on par with other state-of-the-art end-to-end models \cite{CRF_IC19,CAT,CTC-CRF-NAS}.

\section{Multi-channel End-to-end Speech Recognition}
\label{sec:joint}
\begin{figure}[t]

\begin{minipage}[b]{0.45\linewidth}
  \centering
  \centerline{\includegraphics[width=3.0cm]{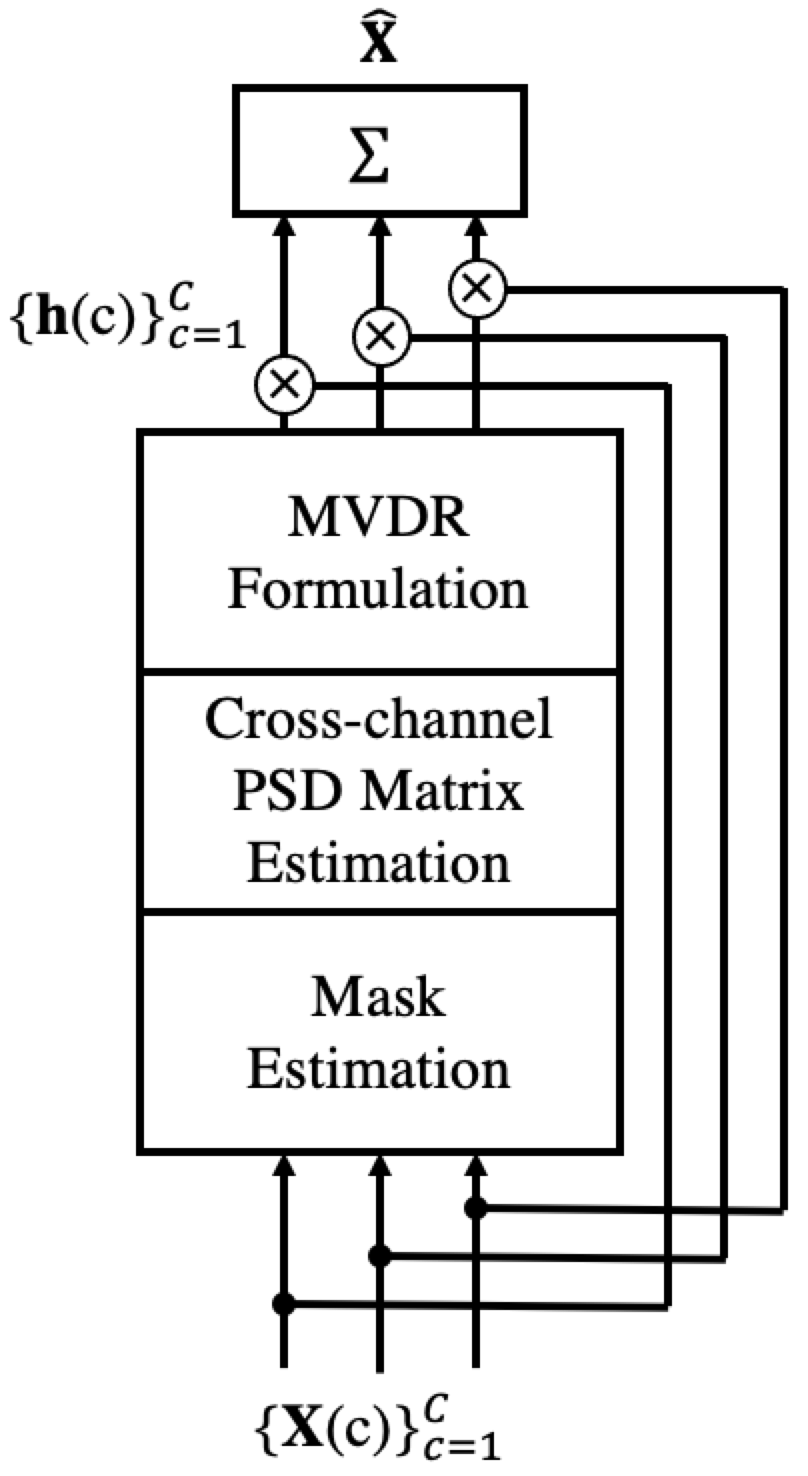}}
  \centerline{(a) MVDR beamformer.} \medskip
\end{minipage}
\hfill
\begin{minipage}[b]{0.45\linewidth}
  \centering
  \centerline{\includegraphics[width=3.0cm]{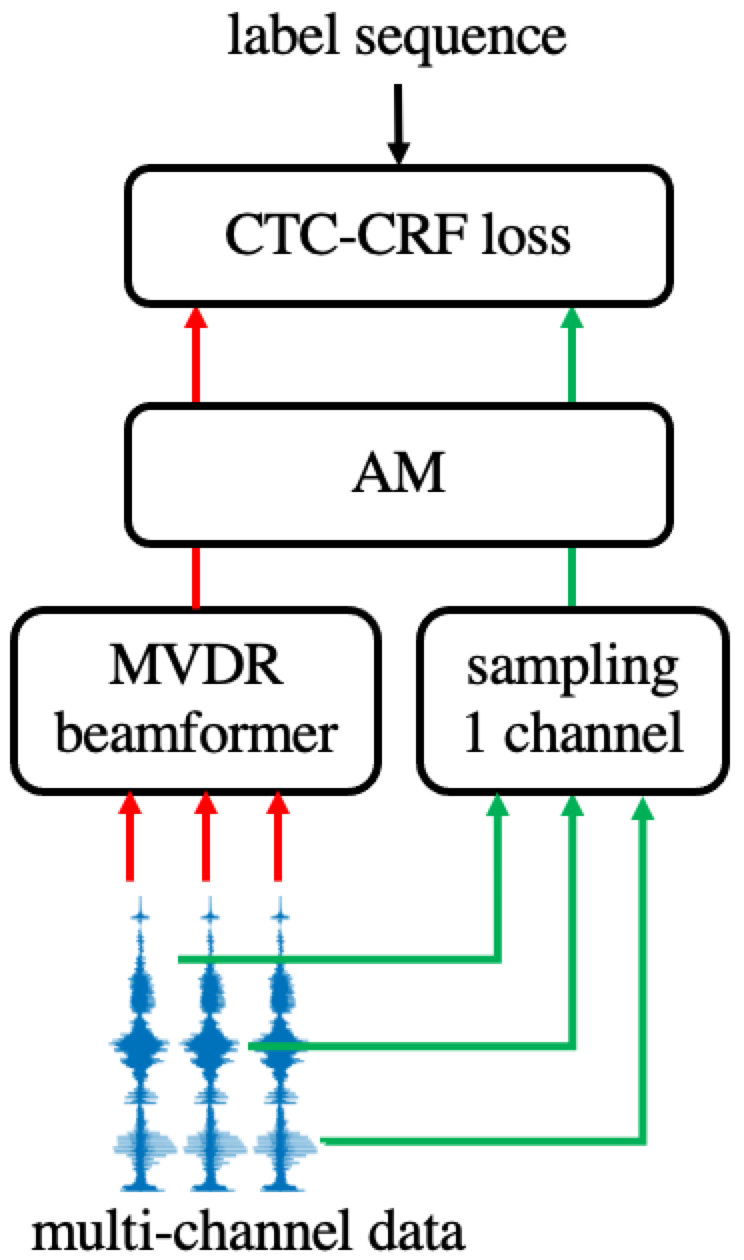}}
  \centerline{(b) Joint training with front-end skipping.
}\medskip
\end{minipage}
\caption{ (a) Illustration of MVDR neural beamformer. (b)  Skipping
the front-end part in the probability of 0.5 when applying front-end and back-end joint training.}
\label{fig:res}
\end{figure}

We adopt a unified architecture for multi-channel end-to-end speech recognition, and apply joint optimization for front-end and back-end. Thus, the training loss is defined as:
\begin{equation} \label{eq:e2e-obj}
\mathcal{L}(\bt) = - \log p_{\bt}(\bm{l}| {\rm Feature} (\hat{\x}))
\end{equation}
where $\hat{\x}$ is obtained by eq (\ref{eq:beforming}), and Feature(·) is a feature extraction function. In this work, we apply log fbank transformation on the STFT coefficients $\hat{\x}$. After that, the features are normalized, appended with delta and delta-delta features, and subsampled to reduce the frame rate. 

According to previous researches \cite{Beamnet_heymann,mimo}, performing purely multi-channel end-to-end training might result in suboptimal results. Following~\cite{mimo}, we skip the front-end part randomly in training (illustrated in Fig. \ref{fig:res}(b)). Specifically, we perform front-end and back-end joint training with probability p (the red flow, Fig. \ref{fig:res}(b)), and select 1 channel from the the multi-channel data and only perform back-end training, with probability 1-p (the green flow, Fig. \ref{fig:res}(b)). With this mechanism, the back-end is trained with both enhanced data and unenhanced noisy data, and would be more robust to the input variations \cite{Beamnet_heymann}. Moreover, the back-end can be independently optimized when the front-end is bypassed, which eases the training. In practice, the probability p is set to 0.5 and fixed. 

\begin{table*}[]
	\centering
	\caption{Effect of joint optimization. The word error rates (WERs) are evaluated on CHiME4.}
	\scalebox{0.9}{
	\begin{tabular}{ccccccc}
		\toprule
		\textbf{Front-end} & \textbf{Back-end}   &  \textbf{Joint optimization} & \textbf{Dev real} & \textbf{Dev simu} & \textbf{Eval real} & \textbf{Eval simu}\\
        \midrule
		BeamformIt~\cite{beamformit} & CTC-CRF & No & 7.28 & 7.98 & 11.11 & 11.97 \\
		MVDR &  CTC-CRF & No & 6.95 & 8.08 & 10.50 & 11.03 \\
		\midrule
		MVDR &  CTC-CRF  & Yes & \textbf{6.15} & \textbf{5.61} & \textbf{9.29} & \textbf{6.14} \\
		\bottomrule
	\end{tabular}}
	\label{joint optimization}
\end{table*}
\begin{table}[]
\caption{Effect of data augmentation and language modeling. The WERs are evaluated on CHiME4. All results from literature adopted neural LM rescoring.}
\begin{threeparttable}
	\centering
	\scalebox{0.85}{
	\begin{tabular}{lcccc}
		\toprule
        \textbf{Model} & \textbf{Dev real} & \textbf{Dev simu} & \textbf{Eval real} & \textbf{Eval simu}\\
        \midrule
		Joint model & 6.15 & 5.61 & 9.29 & 6.14\\
		+ SpecAug & 5.93 & 5.04 & 8.42 & 6.00 \\
		~ + WavAug & 5.60 & 4.94 & 8.06 & 5.70 \\
        ~ ~ + 5gram & 4.52 & 3.87 & 6.42 & 4.41\\
        ~ ~ + RNN LM & 3.66 & 3.17 & 4.80 & 3.52\\
        ~ ~ + LSTM LM & \textbf{2.66} & \textbf{2.05} &  \textbf{3.73}&  \textbf{2.65} \\
        \midrule
        CHiME4 baseline~\cite{chime4} & 5.8 & 6.8 & 11.5 & 10.9 \\
        BEAMNET~\cite{Beamnet_heymann} &  5.51 & 5.19 &  8.76 & 5.61 \\
        USTC-iFlytek\tnote{1}~~\cite{USTC-iFlytek} & \textbf{1.69} & \textbf{1.78} & \textbf{2.24} & \textbf{2.12} \\
		\bottomrule
	\end{tabular}}
	\label{data augmentation}
        \begin{tablenotes}
            \footnotesize
            \item[1] Model ensemble results.
        \end{tablenotes}
\end{threeparttable}
\end{table}

\begin{table*}[]
\caption{Effect of front-end (FE) and back-end (BE) pre-training. The WERs are evaluated on CHiME4.}
\begin{threeparttable}
	\centering
	\scalebox{0.9}{

	\begin{tabular}{cccccccc}

		\toprule
        \textbf{Exp ID} & \textbf{FE pre-training data} & \textbf{BE pre-training data}&  \textbf{Joint optimization} & \textbf{Dev real} & \textbf{Dev simu} & \textbf{Eval real} & \textbf{Eval simu}\\
        \midrule
		1 & No pre-training & No pre-training & yes &5.60 & 4.94 & 8.06 & 5.70\\
        2 & CHiME4 & CHiME4\tnote{1} & no & 6.95 & 8.08 & 10.50 & 11.03 \\
        3 & CHiME4 & CHiME4\tnote{1} & yes & 6.70 & 5.59 & 8.92 & 7.13 \\    
        
        4 & No pre-training & WSJ & yes & 5.84 & 5.16 & 8.05 & 5.78\\
        5 & CHiME4 & WSJ & no & 15.93 & 23.24 & 31.80 & 29.79 \\
        6 & CHiME4 & WSJ & yes & 5.83 & 5.51 & 8.44 & 5.91\\
        
        7 & No pre-training & Librispeech & yes & 4.68 & 4.99 & 7.10 & 5.50\\
        8 & CHiME4 & Librispeech & no & 4.28 & 4.57 & \textbf{5.13} & 6.71\\
        9 & CHiME4 & Librispeech & yes & \textbf{4.22} & \textbf{4.50} & 6.59 & \textbf{4.22}\\
		\bottomrule
	\end{tabular}}
	\label{pre-training}
        \begin{tablenotes}
            \footnotesize
            \item[1] We split CHiME4 into single-channel data and use it to pre-train the back-end.
        \end{tablenotes}
\end{threeparttable}
\end{table*}

\begin{table}[]
	\centering
	\caption{Effect of shared optimizer when applying data scheduling. The single channel data is WSJ. No data scheduling means we don't use any external single-channel data. The WERs are evaluated on CHiME4.}
	\scalebox{0.9}{
	\begin{tabular}{ccccc}
		\toprule
         \textbf{Shared optimizer} & \textbf{Dev real} & \textbf{Dev simu} & \textbf{Eval real} & \textbf{Eval simu}\\
        \midrule
		No data scheduling & 5.60 & 4.94 & 8.06 & 5.70\\
        No  & 5.05 & 4.91 & 7.96 & 5.90 \\
		Yes & \textbf{4.83} & \textbf{4.55} & \textbf{7.17} & \textbf{5.30} \\
		\bottomrule
	\end{tabular}}
	\label{method}
\end{table}

\begin{table*}[]
	\centering
	\caption{Effect of data scheduling with single-channel data in different sizes. The WERs are evaluated on CHiME4.}
	\scalebox{0.9}{
	\begin{tabular}{cccccccc}
		\toprule
        \textbf{Exp ID}&\textbf{Front-end pre-training data} & \textbf{Single-channel data}& \textbf{Dev real} & \textbf{Dev simu} & \textbf{Eval real} & \textbf{Eval simu}\\
        \midrule
		1 & No pre-training & No & 5.60 & 4.94 & 8.06 & 5.70\\
		2 & No pre-training & WSJ & 4.83 & \textbf{4.55} & 7.17 & 5.30\\
        3 & No pre-training & Librispeech & \textbf{4.50} & 4.71 & \textbf{6.30} & \textbf{5.03} \\
        4 & CHiME4 & WSJ & 4.83 & 4.68 & 7.38 & 5.79\\
        5 & CHiME4 & Librispeech & 4.74  & 5.11 & 6.83 & 5.43\\
		\bottomrule
	\end{tabular}}
	\label{mixed training}
\end{table*}
\begin{table}[]
	\centering
	\caption{Effect of training with simulated data. The single channel data for data simulation is WSJ. The WERs are evaluated on CHiME4.}
	\scalebox{0.9}{
	\begin{tabular}{ccccc}
		\toprule
         \textbf{Use simu data} & \textbf{Dev real} & \textbf{Dev simu} & \textbf{Eval real} & \textbf{Eval simu}\\
        \midrule
        No  & 5.60 & 4.94 & 8.06 & 5.70 \\
		Yes & \textbf{5.21} & \textbf{4.41} & \textbf{7.10} & \textbf{4.82} \\
		\bottomrule
	\end{tabular}}
	\label{simulation}
\end{table}
\begin{table}[]
	\centering
	\caption{The character error rate (CER) results on AISHELL-4. The single-channel data for back-end pre-training and data scheduling is AISHELL-1. }
	\scalebox{0.9}{
	\begin{tabular}{cc}
		\toprule
        \textbf{usage of single-channel data} & \textbf{eval CER} \\
        \midrule
		no single-channel data & 60.7 \\
		pre-training & 37.3 \\
		data scheduling &  44.8 \\
		\bottomrule
	\end{tabular}}
	\label{aishell4}
\end{table}
\section{Techniques for exploiting single-channel Speech}
\subsection{Back-end pre-training}
\label{sec:pre}
The first approach to exploit single-channel speech data in multi-channel end-to-end speech recognition system is to do back-end pre-training. Specifically, the training process consists of two stages. In the first stage, the back-end AM is trained with single-channel speech data. In second stage, we perform joint optimization described in Section ~\ref{sec:joint}. In the multi-channel end-to-end speech recognition model, the enhanced log fbank feature produced by the front-end is supposed to be similar to the log fbank of single-channel speech~\cite{mimo}. Thus, back-end pre-training is expected to provide a better initialization for the AM, and lead to less training epochs for joint optimization, which is somewhat more computationally expensive than the single-channel AM training.
\subsection{Data scheduling}
\begin{figure}[ht]
  \centering
  \includegraphics[width=0.45\linewidth]{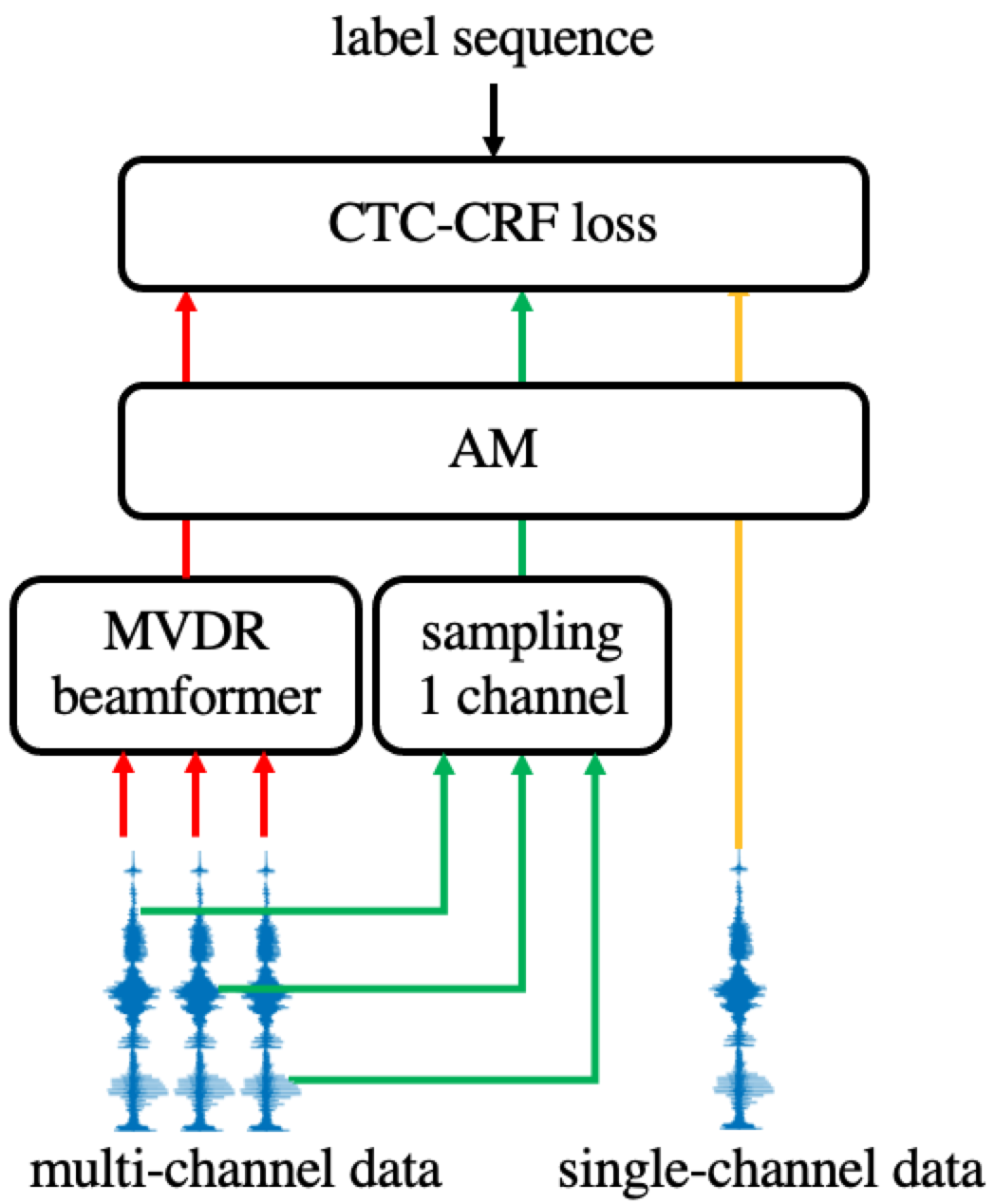}

  \caption{Illustration of data scheduling.}
  \label{fig:mixed}
\end{figure}

The second approach is data scheduling~\cite{mimo}. Different from the two-stage pipeline described in~\ref{sec:pre}, training with data scheduling is conducted in a single stage. As shown in Fig. \ref{fig:mixed}, in data scheduling, the training data comes from two sources: the multi-channel set and the single-channel set. When the training batch comes from the multi-channel set (the red and green flow), we perform joint optimization described in Section ~\ref{sec:joint}. Otherwise, we bypass the front-end and optimize single-channel AM only (the yellow flow), which eases the training like the skip mechanism in Section~\ref{sec:joint}. As the two losses are optimized iteratively, this method allows early stopping by measuring the performance of the multi-channel speech recognition system.  
\subsection{Data simulation}
The third approach is to simulate multi-channel data using single-channel data~\cite{pyroomacoustics}. In this approach, we first define a room to which the sound source (the single-channel wave samples) and a microphone array are attached. Then, a simulation method is used to create artificial room impulse responses (RIRs) between the source and microphones. The microphone signals are then created by convolving the single-channel wave samples with the  RIRs. After data simulation, the simulated multi-channel data are mixed with the real multi-channel data to train the multi-channel speech recognition system. In conventional systems combined with a separate front-end and back-end, it has been shown that the beamforming module can benefit from training on simulated data~\cite{chime4}. However, the mismatch between the enhanced real and simulated data may negatively affect the back-end training~\cite{chime4}. Another obvious drawback is that the computational complexity of training with simulated data is higher than pre-training and data scheduling with the same amount of single-channel data. 
\section{Experiments}
\subsection{Datasets}
\subsubsection{Multi-channel Datasets}
\begin{itemize}
\item CHiME4~\cite{chime4}  is a speech recognition task in public noisy environments, recorded using a 6-channel tablet based microphone array. The corpus is in English and the training data length is 18 hours. The test set is made of 4 subsets: Dev Real, Dev Simu,	Eval Real, and Eval Simu.
\item AISHELL-4~\cite{aishell4} is a multi-channel mandarin dataset for conversation speech in conference scenarios, containing 118 hours of meeting recording, recorded using an 8-channel microphone array. As the integration of the speaker diarization module is beyond the scope of this paper, we select the non-overlapped part of the training and evaluation set of AISHELL-4 according to the ground-truth segmentation information. For running the AISHELL-4 experiments, we use the open-source lexicon provided by AISHELL-1 dataset \footnote{{http://www.openslr.org/resources/33/resource\_aishell.tgz}} as there is no official lexicon in the AISHELL-4 dataset, and we perform transcript pre-processing using the Jieba segmentation toolkit \footnote{{https://github.com/fxsjy/jieba}} for word segmentation.
\end{itemize}
\subsubsection{Single-channel Datasets}
\begin{itemize}
\item WSJ~\cite{wsj} contains about 80 hours of training data recorded under clean conditions. The corpus is consists of read English sentences from the Wall Street Journal.
\item Librispeech~\cite{librispeech} contains 1000 hours of English read speech, derived from audiobooks.
\item AISHELL-1~\cite{aishell} is a 178-hour mandarin speech corpus.
\end{itemize}
\subsection{Experiment settings}
We use the CTC-CRF based ASR Toolkit - CAT~\cite{CAT} to conduct the experiments.
In our experiment, the inputs to the front-end are STFT features. The mask estimation network in the neural beamformer is a 3-layer BLSTM. After beamforming, the enhanced single-channel STFT features are firstly converted to 40-dimensional log fbank features, and then mean-variance normalized. The normalized log fbank features are appended with delta and delta-delta features and subsampled by a factor of 3. Similar to~\cite{CAT}, the acoustic model is two blocks of
VGG layers followed by a 6-layer BLSTM, and the BLSTM
has 320 hidden units per direction.  During training, a dropout probability of 50\% is applied to the BLSTM to prevent overfitting. We choose Adam as the optimizer, and apply gradient clipping to avoid the training failure caused by unexpected exploding gradients.

In data simulation experiments, we adopt pyroomacoustics~\cite{pyroomacoustics} to simulate multi-channel waves using single-channel speech as the source signal.  We define a 10m $\times$ 7.5m $\times$ 3.5m room, and the source is located at [2.5, 3.73, 1.76]. The microphone configuration is the same as the one used in CHiME4 challenge \footnote{{http://spandh.dcs.shef.ac.uk/chime\_challenge/CHiME4/overview.html}}, which is a 6-channel microphone array embedded in the frame of a tablet device: 3 along the top and 3 along the bottom. In our experiment, the center of the microphone is fixed at [5, 2.25, 1] for simplicity.

Note that we did not use the transcripts of single-channel data in language model building. In CHiME4 experiments, we use the ngram and neural language model provided by the challenge. In AISHELL-4 experiments, we use a 3-gram language model trained on the AISHELL-4 training set transcripts.
\subsection{Joint optimization}
The effect of joint optimization is shown in Table~\ref{joint optimization}. We compare the performance of the jointly optimized model with two baseline models, in which the enhanced speech data produced by the delay-and-sum beamformer (BeamformIt) or pre-trained neural beamformer are used as input to the back-end.
It can be seen that the jointly optimized model significantly outperforms the two baseline models, which confirms the effectiveness of combining speech enhancement with the ASR module.

\subsection{Data augmentation and language modeling}
We augment the input to the front-end and back-end independently. Specifically, we apply WavAugment~\cite{wavaug} on the input audio, and SpecAugment~\cite{specaug} on the log fbank features. The effect of data augmentation is shown in Table~\ref{data augmentation}. It can be seen that the combination of WavAugment and SpecAugment yields the best performance. Therefore we adopt this technique in the rest of the experiments. 

Then we evaluate the model performance with respect to how they are affected by language models. Specifically, we decode the test data using a 3-gram language model, and then rescoring the results using a 5-gram, an RNN, and finally an LSTM language model.  It can be seen that the WER drops dramatically as stronger language models are used.

We also compare our model with models from literature in Table~\ref{data augmentation}. It can be seen that our model outperforms the CHiME4 baseline system~\cite{chime4} significantly. Compared with BEAMNET~\cite{Beamnet_heymann}, which performs front-end and back-end joint optimization from scratch as well, our model achieves better recognition performance. Compared with CHiME4 USTC-iFlytek system~\cite{USTC-iFlytek}, which is the rank 1 system in CHiME4 challenge, our model performs comparably, and it is worth pointing out we do not use model ensemble as in~\cite{USTC-iFlytek}.

In the following sections, we will evaluate the effectiveness of exploiting external single-channel speech to further improve the multi-channel end-to-end speech recognition accuracy. For simplicity, we use the 3-gram language model for decoding in the following experiments.

\subsection{Pre-training}
The effect of pre-training is shown in Table~\ref{pre-training}. In addition to back-end pre-training using single-channel speech, we also explored front-end pre-training using the parallel clean and noisy provided by CHiME4. From Table~\ref{pre-training}, we find that: 
1) joint optimization after pre-training is essential to improve the multi-channel ASR performance (Exp 2 vs Exp 3, Exp 5 vs Exp 6, Exp 8 vs Exp 9). 
2) Back-end pre-training on CHiME4 and WSJ shows no significant improvement in the recognition accuracy (Exp 2 $\sim$ Exp 6), which is presumably due to the limited data size. Notably, the performance of pre-training on WSJ without joint optimization (Exp 5) is quite poor, and we assume that this is because the back-end overfits the WSJ data. 
3) Back-end pre-training on 1000-hour Librispeech produces substantial improvement (Exp 7 $\sim$ Exp 9). Front-end pre-training also improves the performance (Exp 7 vs Exp 9). 4) When the back-end pre-training data is increased, the performance after joint optimization becomes better (Exp 3 vs Exp 6 vs Exp 9). 

It is worth noting that joint optimization from a pre-trained front-end and back-end leads to a much faster convergence compared with joint optimization from scratch. In our experiments, training from scratch typically takes more than 20 epochs to converge, while training from a pre-trained model usually takes less than 10 epochs to converge. Moreover, initialization with a pre-trained model leads to a much more stable training process. The optimization works well even without the use of techniques like gradient clipping. 

\subsection{Data scheduling}
In this section, we analyze how data scheduling improves the multi-channel end-to-end speech recognition system.
First, we compare two training methods when applying data scheduling. The first is to use two separate optimizers for the multi-channel data flow and the single-channel data flow, which means the two optimizers maintain their state independently while training. The second is to use a single optimizer: the optimizer is shared when the training batches come from two sources. The result in Table~\ref{method} shows that data scheduling with a single optimizer yields better performance. Therefore we adopt this setting in the rest of the experiments with data scheduling. 
Table~\ref{mixed training} presents the results of data scheduling with single-channel data in different sizes. From Table~\ref{mixed training}, we observe that 1) When the back-end is not pre-trained, using a pre-trained front-end seems to have no significant effect on the recognition accuracy. (Exp 2 vs Exp 4, Exp 3 vs Exp5); 2) When the single-channel data for data scheduling is increased, the performance of the multi-channel system gets better (Exp 2 vs Exp 3, Exp 4 vs Exp 5).

Similar to pre-training, we empirically find that the use of single-channel data stabilizes the training process, as it takes fewer training epochs to converge and rarely falls into the training failure (e.g., the training loss divergence) even without the use of gradient clipping.
\subsection{Data simulation}
The effect of training a multi-channel end-to-end system using simulated data is shown in Table~\ref{simulation}. Consistent and substantial improvement is observed on the test sets, and the performance is comparable with pre-training and data scheduling. Nevertheless, the time cost of using simulated data is much higher than the two fore-mentioned methods, as the memory and computation cost of multi-channel end-to-end training is more expensive than the single-channel AM training. To be specific, simulating and training with WSJ takes approximately 2$\times$ overall time as data scheduling with WSJ.

\subsection{Results on AISHELL-4}
The results of  AISHELL-4 are shown in Table~\ref{aishell4}. We did not perform training using simulated data due to the large time and computation overhead. Similar to the findings on CHiME4, exploiting single-channel data leads to large improvements in the recognition accuracy of multi-channel end-to-end speech recognition system.

\section{Conclusion}
In this paper, we explore three simple yet effective methods to exploit single-channel data in a multi-channel end-to-end speech recognition system. Extensive experiments and thorough analysis demonstrate that the use of single-channel data improves the end-to-end multi-channel system in training stability and recognition accuracy.

\bibliographystyle{IEEEbib}
\bibliography{Template_Regular}

\end{document}